\documentclass[sigconf, nonacm]{acmart}

\usepackage{enumitem}
\usepackage{float}

\AtBeginDocument{%
  }

\copyrightyear{2025}
\acmYear{2025}
\setcopyright{cc}
\setcctype{by}
\acmConference[SIGIR '25]{Proceedings of the 48th International ACM SIGIR Conference on Research and Development in Information Retrieval}{July 13--18, 2025}{Padua, Italy}
\acmBooktitle{Proceedings of the 48th International ACM SIGIR Conference on Research and Development in Information Retrieval (SIGIR '25), July 13--18, 2025, Padua, Italy}
\acmDOI{10.1145/3726302.3730327}
\acmISBN{979-8-4007-1592-1/2025/07}

\begin{document}
\pagestyle{plain}

\title{Interpreting Multilingual and Document-Length Sensitive Relevance Computations in Neural Retrieval Models through Axiomatic Causal Interventions}


\author{Oliver Savolainen}
\authornote{These authors contributed equally to the paper.}
\affiliation{%
  \institution{University of Amsterdam}
  \city{Amsterdam}
  \country{Netherlands}}
\email{oliver.savolainen@student.uva.nl}

\author{Dur e Najaf Amjad}
\authornotemark[1]
\affiliation{%
  \institution{University of Amsterdam}
  \city{Amsterdam}
  \country{Netherlands}}
\email{najaf.amjad@student.uva.nl}

\author{Roxana Petcu}
\affiliation{%
  \institution{University of Amsterdam}
  \city{Amsterdam}
  \country{Netherlands}}
\email{r.m.petcu@uva.nl}


\renewcommand{\shortauthors}{}

\begin{abstract}
This reproducibility study analyzes and extends the paper "Axiomatic Causal Interventions for Reverse Engineering Relevance Computation in Neural Retrieval Models," which investigates how neural retrieval models encode task-relevant properties such as term frequency. We reproduce key experiments from the original paper, confirming that information on query terms is captured in the model encoding. We extend this work by applying activation patching to Spanish and Chinese datasets and by exploring whether document-length information is encoded in the model as well. Our results confirm that the designed activation patching method can isolate the behavior to specific components and tokens in neural retrieval models. Moreover, our findings indicate that the location of term frequency generalizes across languages and that in later layers, the information for sequence-level tasks is represented in the CLS token. The results highlight the need for further research into interpretability in information retrieval and reproducibility in machine learning research. Our code is available at \url{https://github.com/OliverSavolainen/axiomatic-ir-reproduce}.
\end{abstract}




\maketitle

\section{Introduction}
\label{sec:introduction}

Transformers have achieved state-of-the-art performance across various domains, including information retrieval. Their inner workings have been extensively studied, to make their reasoning process more transparent and trustworthy. The paper "Axiomatic Causal Interventions for Reverse Engineering Relevance Computation in Neural Retrieval Models" \cite{orig} explores an approach to understanding how a neural retrieval model makes its modelling decisions. 

\textbf{Original findings}. \citet{orig} apply activation patching to TAS-B \cite{TASB}, a BERT-based neural retrieval model. The activation patching method uses a carefully crafted diagnostic dataset that tries to isolate term frequency information through synthetic query-document pairs \cite{axioms}. This work presents the following key findings:

\textbf{1)} Activation patching can effectively isolate a task-dependent behaviour to specific components and tokens in neural retrieval models.

\textbf{2)} Term frequency information is stored across the occurrences of the query term, with a focus on the first term positions in the document, in the first layers of the model. Later, this information is stored in the pooled representation of the CLS token.

\textbf{3)} The attention mechanism includes four attention heads specific for processing term frequency information. These heads work as two different purpose groups in the initial layers and middle layers that interact with each other. \cite{orig}

\textbf{Our findings}. To validate the claims made, we aim to reproduce and further extend the paper by \citet{orig}. The first extension examines the inner workings of the model across different languages, specifically Spanish and Chinese. The other extension uses the LNC1 axiom to investigate whether the model encodes information on document length, a critical feature for effective ranking models. Our experiments confirm that the model stores term frequency information across different languages, with an emphasis on the first positions of the document. We also show that the model stores document length information at the CLS token. While we think that the designed method can isolate the behaviour of specific components of neural retrieval models, it could be backed up by other interpretability methods such as circuit analysis and sparse autoencoders \cite{ap, sae}.

The paper is organized as follows. In Section 2, we review the relevant literature. Section 3 outlines the methodology used in the original study. In Section 4, we introduce two extensions to the original approach. In Section 5, we describe the experimental setup and in Section 6 the results of both the reproduced and the extension experiments are discussed. Section 7 consists of discussing the original paper and findings from our work. Finally, conclusions are made in Section 8.
\section{Related Work}
\label{sec:related_work} 

\textbf{Activation Patching.} Causal interventions differentiate themselves from other interpretability methods, such as probing \cite{probing, belinkov2021probingclassifierspromisesshortcomings}, by providing causal insights into the model behaviour. One such method is activation patching, for which a corrupted input $X_{\text{perturbed}}$ with its output $Y_{\text{perturbed}}$ is needed alongside a clean input $X_{\text{baseline}}$ with the correct output $Y_{\text{baseline}}$. For example, $X_{\text{baseline}}$ could be “Paris is the capital of”, with $Y_{\text{baseline}}$ “France” while $X_{\text{perturbed}}$ is “London is the capital of” and the expected answer $Y_{\text{corrupt}}$ for it is "England." To perform activation patching, the model is first run with $X_{\text{baseline}}$, and its activations are cached. The model is then run with $X_{\text{perturbed}}$, and selected activations are replaced with those from the run with $X_{\text{baseline}}$. To understand the impact of specific components from the model, we measure how well we can recreate the result of $X_{\text{baseline}}$ by iteratively replacing activations. In the example, it would be about how close we reach to getting $Y_{\text{baseline}}$ “France” \cite{orig, zhang2024bestpracticesactivationpatching,parry2025mechirmechanisticinterpretabilityframework}.

\textbf{TFC1 and LNC1 Axioms}.
Axiomatic Information Retrieval (IR) has focused on defining retrieval principles to guide neural ranking models and their evaluation.
The TFC1 axiom \cite{axioms} examines the relationship between term frequency and retrieval effectiveness, emphasizing that higher query term frequency in a document should lead to higher document relevance scores, provided all other factors remain equal.
The LNC1 axiom \cite{LNC1} states that if two documents contain the same terms with the same term frequencies, but one document is longer due to additional non-relevant terms, the relevance score of the longer document should not be higher than the shorter one. In a more recent study, LNC1 is analyzed in relation to its agreement with key retrieval axioms, particularly those addressing term frequency saturation and document length normalization \cite{Thakur_2024}.

\textbf{mMARCO}. The mMARCO dataset \cite{bonifacio2022mmarcomultilingualversionms} plays a crucial role in addressing the gap of multilingual open-source information by overcoming the bias toward English-centric models and enabling a more robust evaluation of retrieval across diverse languages. As a multilingual benchmark dataset, mMARCO allows models such as ListConRanker \cite{liu2025listconrankercontrastivetextreranker} to leverage its extensive and diverse passages, facilitating global contrastive information learning and interaction within the ListTransformer. 
\section{Methodology}
\label{sec:methodology}

The methodology consists of 3 parts: creating the diagnostic dataset, running the activation patching experiments, and analyzing the results as internal mechanisms of the model \cite{orig}.

\subsection{Diagnostic Dataset}

\begin{figure*}
    \centering
    \includegraphics[width=0.75\linewidth]{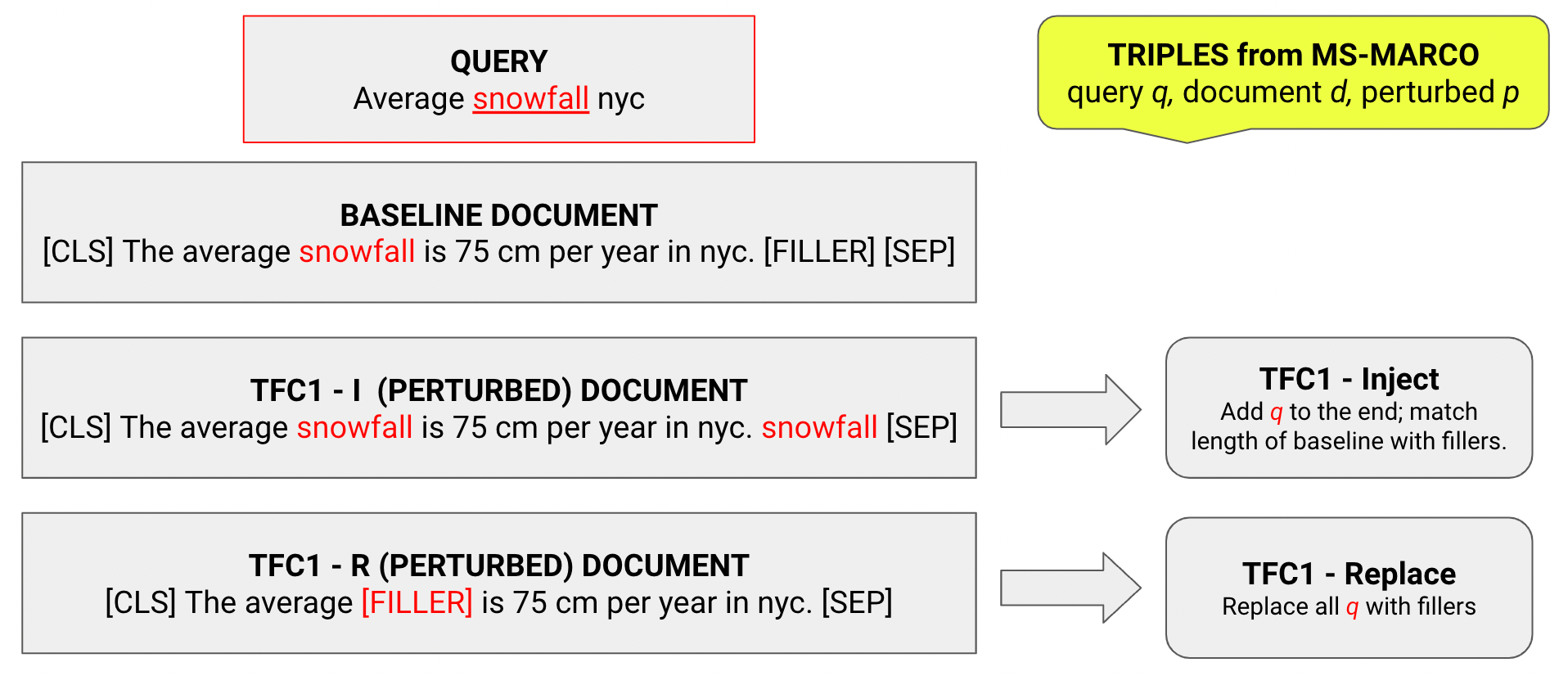}
    \caption{Diagnostic dataset creation setup for TFC1-I and TFC1-R.}
    \label{fig:diagnostic_dataset}
\end{figure*}

\citet{orig} study the TFC1 \cite{axioms} axiom to analyze the term frequency information of query terms present in the documents. To perform activation patching, a diagnostic dataset is designed by creating $(q, d_b, d_p)$ triples from the MS-MARCO developmental dataset \cite{msm} where one of the documents is in its original form, and the other is perturbed based on a randomly sampled query term. Next, the relevance scores from the model outputs are compared between the perturbed and the patched input runs to measure the impact of specific components \cite{orig}. The dataset is created based on the TFC1 axiom which is defined by \citet{orig} as: 

\textbf{TFC1}. Let $q = w$ be a query with only one term $w$. Assume the length of document $d_1$ equals the length of document $d_2$. If the number of occurrences of $w$ in $d_1$ is greater than the number of occurrences of $w$ in $d_2$, then for query $q$, the relevance score of $d_1$ should be higher than $d_2$.

\vspace{1em}

\citet{orig} define two perturbations to observe the effects of TFC1: injection and replacement. TFC1-Inject (TFC1-I) and TFC1-Replace (TFC1-R) are outlined below:

\begin{description}
  \item[\textbf{TFC1-I.}]
  We randomly sample a query term and insert it at the end of the document \(d\) 
  to create our perturbed document \(d_p\). To create a baseline document \(d_b\) 
  equal in length to our perturbed document, we insert a filler token 
  (e.g., `a') at the end of document \(d\).

  \item[\textbf{TFC1-R.}]
  We randomly sample one query term and replace all its occurrences in document 
  \(d\) with a filler token to create a perturbed document \(d_p\). The original 
  document \(d\) acts as the baseline document \(d_b\).
\end{description}

The dataset creation process can be seen in Figure \ref{fig:diagnostic_dataset}. For TFC1-I, a randomly sampled query term is added either at the start, called prepend, or at the end of the document, called append. This document will now be the perturbed input. The other document in the triple is the baseline document, i.e. the original document, with filler tokens 'a' injected in the place of the added query term in the perturbed document. For TFC1-R, every appearance of the sampled query term is replaced by the filler token and the baseline document is just the original document. For each query, the 100 most relevant documents are retrieved and perturbed. For the final dataset, the 100 queries with the biggest average change of relevance score after perturbation were identified \cite{orig}. For our reproducibility experiments, we used the final version of the dataset given to us instead of reproducing the dataset due to the lack of details in the paper.

\subsection{Activation Patching in Retrieval}

\begin{figure*}[ht]
    \centering
    \includegraphics[width=1\linewidth]{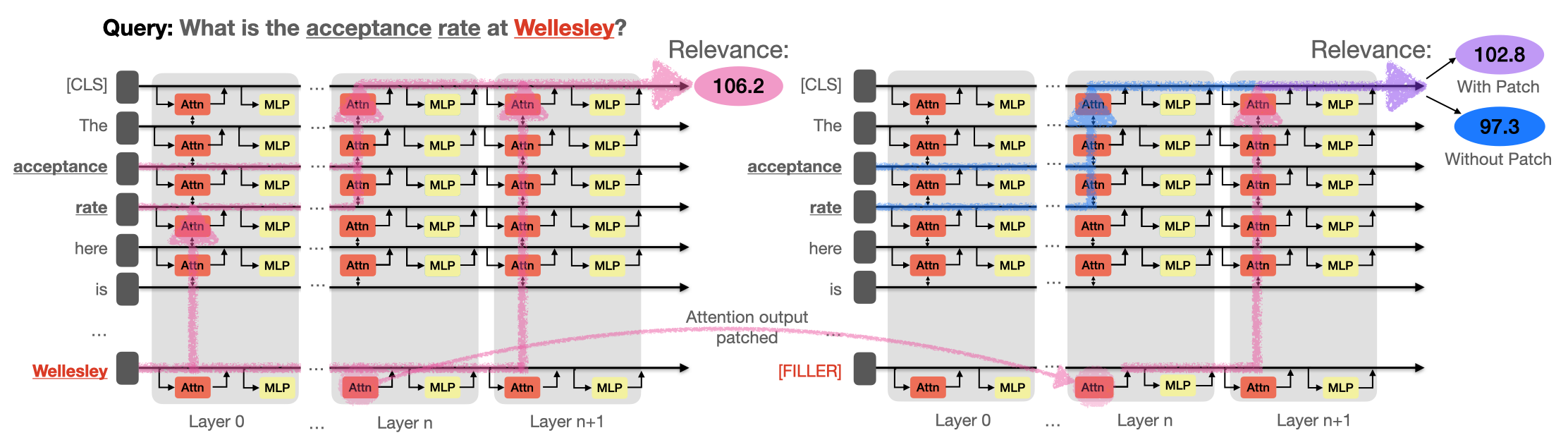}
    \caption{Activation patching setup for retrieval from the original paper. In this example, a $(q, d_b, d_p)$ triple is created to measure term frequency effects in the model. A perturbed document (left) is created by injecting the sampled query term (“Wellesley”) at the end, and a baseline document (right) is made by adding filler tokens. When running the model with the perturbed input, the activations are saved. For the patched run, the model runs on the baseline document, and the cached activations are patched in. This should lead to an increase in the ranking score. \cite{orig}}
    \label{fig:patching}
\end{figure*}

The $(q, d_b, d_p)$ triples from the diagnostic dataset are used to apply activation patching for the neural retrieval model. As seen in Figure \ref{fig:patching}, the patched run has $X_{\text{baseline}}$ as input with cached activations coming from running the model with $X_{\text{perturbed}}$ as perturbations should lead to increases in relevance. This is the opposite for TFC1-R as that perturbation leads to a decrease in relevance \citet{orig}. Since we are working with a retrieval model, the dot product of query and document embeddings is used as a metric. The formula for measuring the difference between scores before and after patching is defined by \citet{orig} as:

\begin{equation}
\text{Patching Impact} = 
\frac{\text{Score}_{X_{\text{patched}}} - \text{Score}_{X_{\text{baseline}}}}{\text{Score}_{X_{\text{perturbed}}} - \text{Score}_{X_{\text{baseline}}}}
\end{equation}

This formula measures how much (\(X_{\text{patched}}\)) shifts away from (\(X_{\text{baseline}}\)) and towards the relevance score of (\(X_{\text{perturbed}}\)). For TFC1-I, we are expecting the relevance score to increase after the perturbation. Therefore, the higher the metric value, the more important the activation is to the result. Based on the axiom, the patching should show how and where the term frequency information of the selected query term is stored.

\section{Extension}

We propose two extensions to validate the results from the original paper and further analyze the TAS-B model's behavior. First, we perform causal interventions based on the LNC1 axiom to understand how the model encodes document-length information. Secondly, we use the mMARCO Spanish and Chinese language datasets to understand how the term frequency localisation compares between different languages.

\subsection{LNC1}

To better understand the internal mechanisms of the retrieval model, we apply axiomatic causal interventions based on another axiom, namely LNC1. The axiom can be described as the following \cite{axioms}:

\textbf{LNC1}. Let $q$ be a query and $d_1, d_2$ be two documents. Assume that for some word $w' \notin q$, $c(w', d_2) = c(w', d_1) + 1$. If for any query term $w$, $c(w, d_2) = c(w, d_1)$, then for query $q$, the relevance score of $d_1$ should be greater than or equal to $d_2$. 

This principle can be used to evaluate how effectively neural retrieval models handle document-length variability, a critical aspect in datasets like Touché 2020 \cite{t2020}, which often contain noisy, short, and non-argumentative documents. While traditional models such as BM25 show near-perfect adherence to LNC1, neural models like TAS-B \cite{TASB} and SPLADEv2 \cite{formal2021spladev2sparselexical} frequently violate this axiom, leading to a suboptimal ranking of documents with varying lengths. This analysis reveals that LNC1 serves as a diagnostic tool to pinpoint specific weaknesses in neural retrieval systems, such as their inability to prioritize longer, more detailed arguments over shorter, less informative ones. 

We are using a similar experimental setup as the original paper, with a few modifications to analyze the length property. The variants of prepending and appending query terms are not relevant for the LNC1 experiments. Instead, we add terms that are not relevant to the query to decrease relevancy. At first, we experimented with adding a single filler token. However, that did not lead to a consistent enough decrease in the relevance score. The same behaviour has been observed with the use of a token not present in the queries, such as "guantanamo". This eventually leads us to doubling the document size with the injected tokens.

\begin{figure}
    \centering
    \includegraphics[width=1\linewidth]{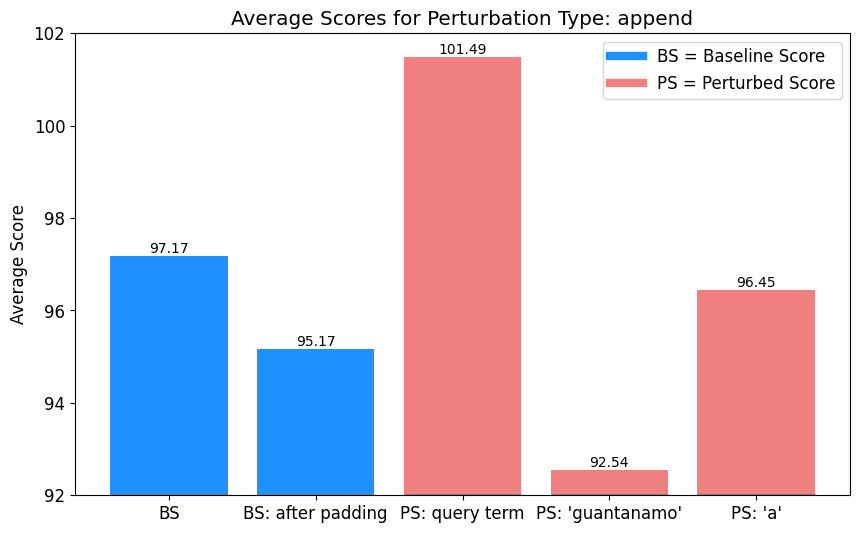}
    \caption{The average relevance score comparison shows the baseline score with and without padding. The perturbed conditions include appending a query term, 'guantanamo' tokens, and 'a' tokens. }
    \label{fig:scores}
\end{figure}

Moreover, both inputs must have the same length. To achieve this, we tested masking the injected tokens, which resulted in a clear decrease in relevance. As seen in Figure \ref{fig:scores}, the most consistent way of achieving a decrease in relevance after the perturbation is to use padding tokens for $X_{\text{baseline}}$ and 'guantanamo' tokens for $X_{\text{perturbed}}$, as 'guantanamo' doesn't appear a single time in the documents and queries in the final dataset.

The experiment involves inserting 'guantanamo' tokens to construct $X_{\text{perturbed}}$, and padding tokens in place of them for $X_{\text{baseline}}$. Next, activations from the perturbed run are saved and used to patch activations when running the model with the padded baseline document. This approach differs from the original paper, as their saved activations are typically from the higher-relevance run. However, given the design of the ranking metric used, the patched activations should still highlight differences in length information.

\subsection{mMARCO Dataset - Spanish and Chinese}

In this extension, we want to determine whether the conclusions drawn from the previous experiment on an English-language dataset hold across different languages. To achieve this, we use the mMARCO dataset \cite{bonifacio2022mmarcomultilingualversionms}, the multilingual version of MS MARCO, which includes 13 languages, excluding the original English version. We experiment with Spanish and Chinese languages to see if fundamental differences between languages impact the behaviour of the model.

The experimental setup is similar to MS MARCO in the original paper. To find the top 100 relevant documents, we only use documents from the corpus associated with queries instead of going through the entire corpus. For the final dataset, we chose the 100 queries with the highest relative change from the append perturbation as \citet{orig} did not specify which perturbation was used to select the queries. 

\label{sec:extensions}

\section{Experimental Setup}
\label{sec:experimental_setup}

The model used in the original work is TAS-B \cite{TASB}, a DistilBERT-based model with 6 layers and 12 attention heads per attention layer. TAS-B is a bi-encoder, which processes queries and documents separately and uses a pooled representation of the CLS token for ranking score calculation. The choice of the model was due to its high performance in ranking tasks despite its simplified architecture. \cite{orig}

Activation patching is done in multiple parts of the transformer: (1) residual stream, (2) attention outputs, (3) MLP outputs, and (4) individual attention heads. The first 3 are done by applying patching for each token in the document, which is referred to as the block experiment. Most of the results are based on the TFC1-I appending perturbation \cite{orig}.

The patching experiments quantify the impact of each component in terms of term frequency information. The tokens are classified as per Table \ref{tab:token_classifications}. The darker the shade of blue in the square of the heatmap, the more significant the impact of the activations at those tokens is in increasing performance.

\begin{table}[h!]
\centering
\caption{Token type classifications for documents. TFC1-I perturbed documents include all six token types, while TFC-R perturbed documents have five token types since no terms are injected during perturbation.}
\begin{tabular}{|c|p{6.5cm}|}
\hline
\textbf{Label} & \textbf{Definition} \\ \hline
\texttt{tok\textsubscript{CLS}} & The CLS token. \\ \hline
\texttt{tok\textsubscript{inj}} & The selected query term injected into the document. \\ \hline
\texttt{tok\textsubscript{qterm\textsuperscript{+}}} & Occurrences of the selected query term that already exist in the original document. \\ \hline
\texttt{tok\textsubscript{qterm\textsuperscript{--}}} & Occurrences of the non-selected query terms in the original document. \\ \hline
\texttt{tok\textsubscript{other}} & Terms in the original document that are not query terms. \\ \hline
\texttt{tok\textsubscript{SEP}} & The SEP token. \\ \hline
\end{tabular}
\label{tab:token_classifications}
\end{table}
\section{Experimental Results}
\label{sec:experimental_results}

This section describes the results of reproducing the original paper and the two extensions we designed. Similar to the experiments in \cite{orig}, we report results from a single run. A fixed random seed is set, and we use a batch size of 1.

\subsection{Reproduced Results}

\subsubsection{Importance of Added/Deleted Query Terms (The Block Experiment)} 
This experiment investigates how specific activations within the residual stream and attention outputs can help identify the tokens that have the most significant impact on performance.

\begin{figure}
    \centering
    \includegraphics[width=1\linewidth]{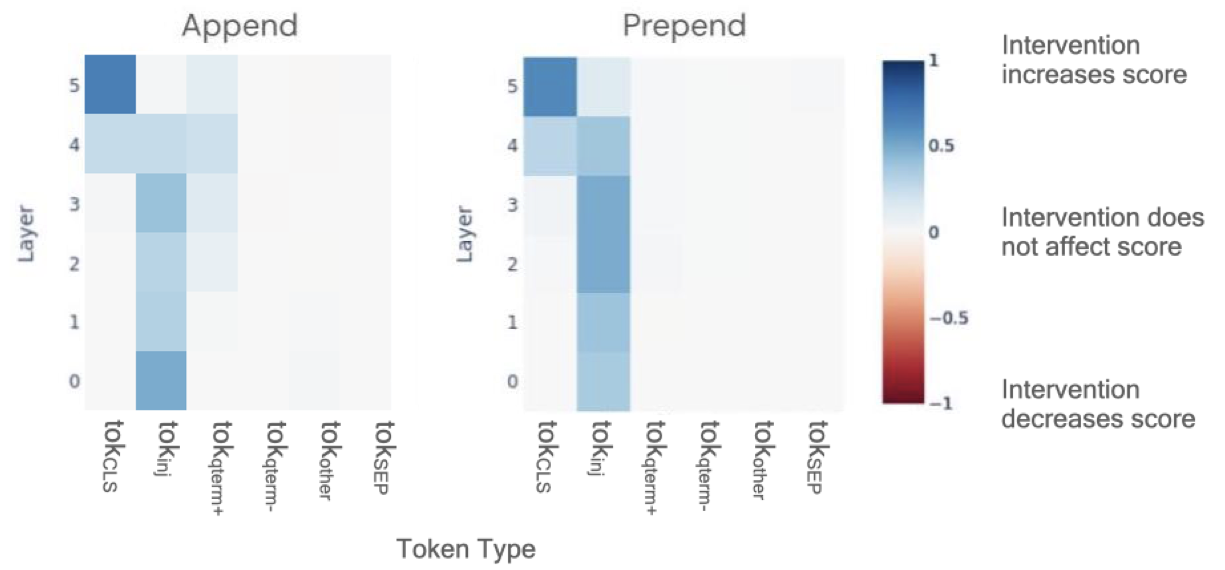}
    \caption{The original block experiment with append and prepend from patching into the residual stream.}
    \label{fig:block-append-prepend-original}
\end{figure}

\begin{figure}
    \centering
    \includegraphics[width=1\linewidth]{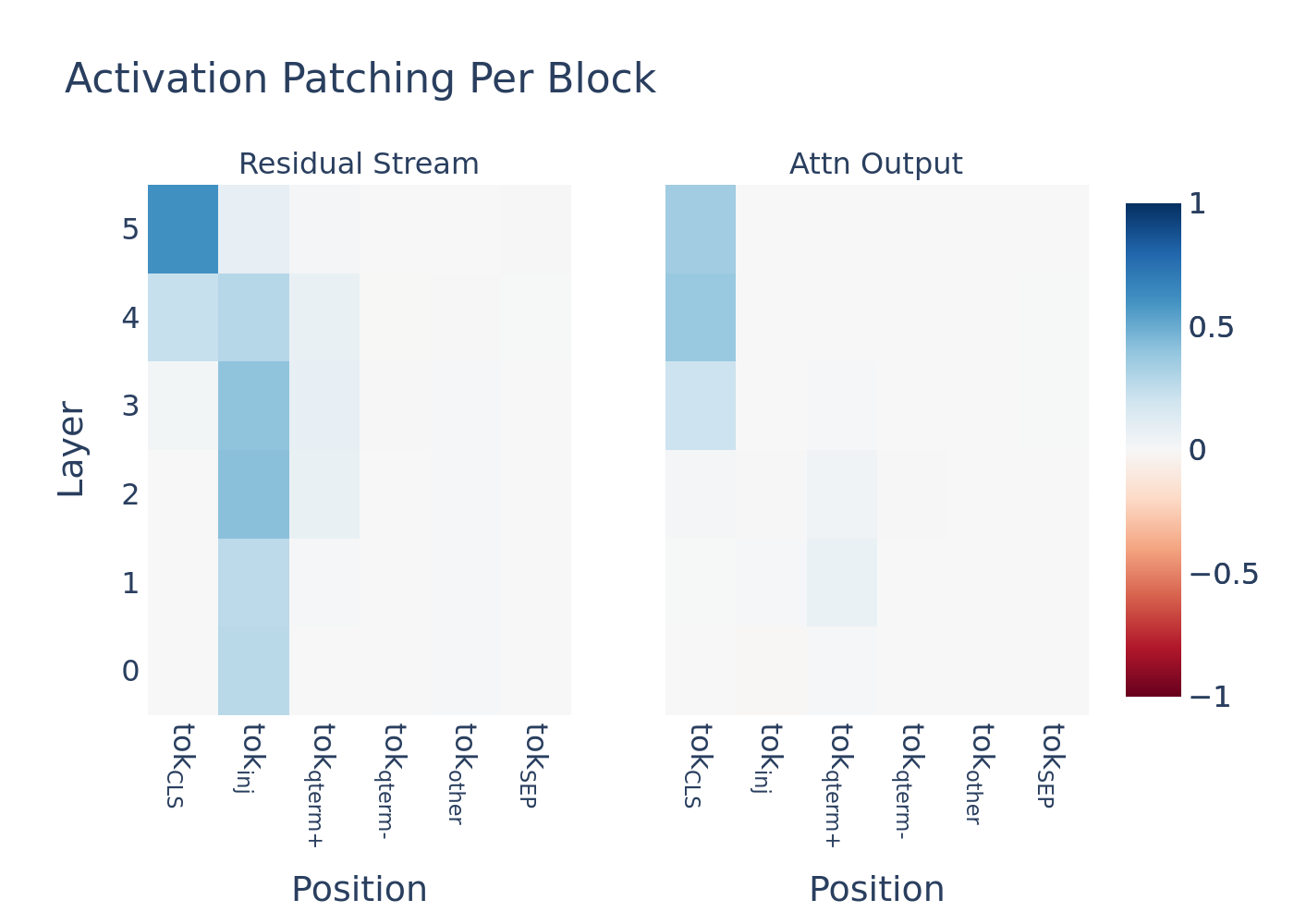}
    \caption{The reproduced block experiment results for append from patching into the residual stream and attention output.}
    \label{fig:append_all_block_seg}
\end{figure}

\begin{figure}
    \centering
    \includegraphics[width=1\linewidth]{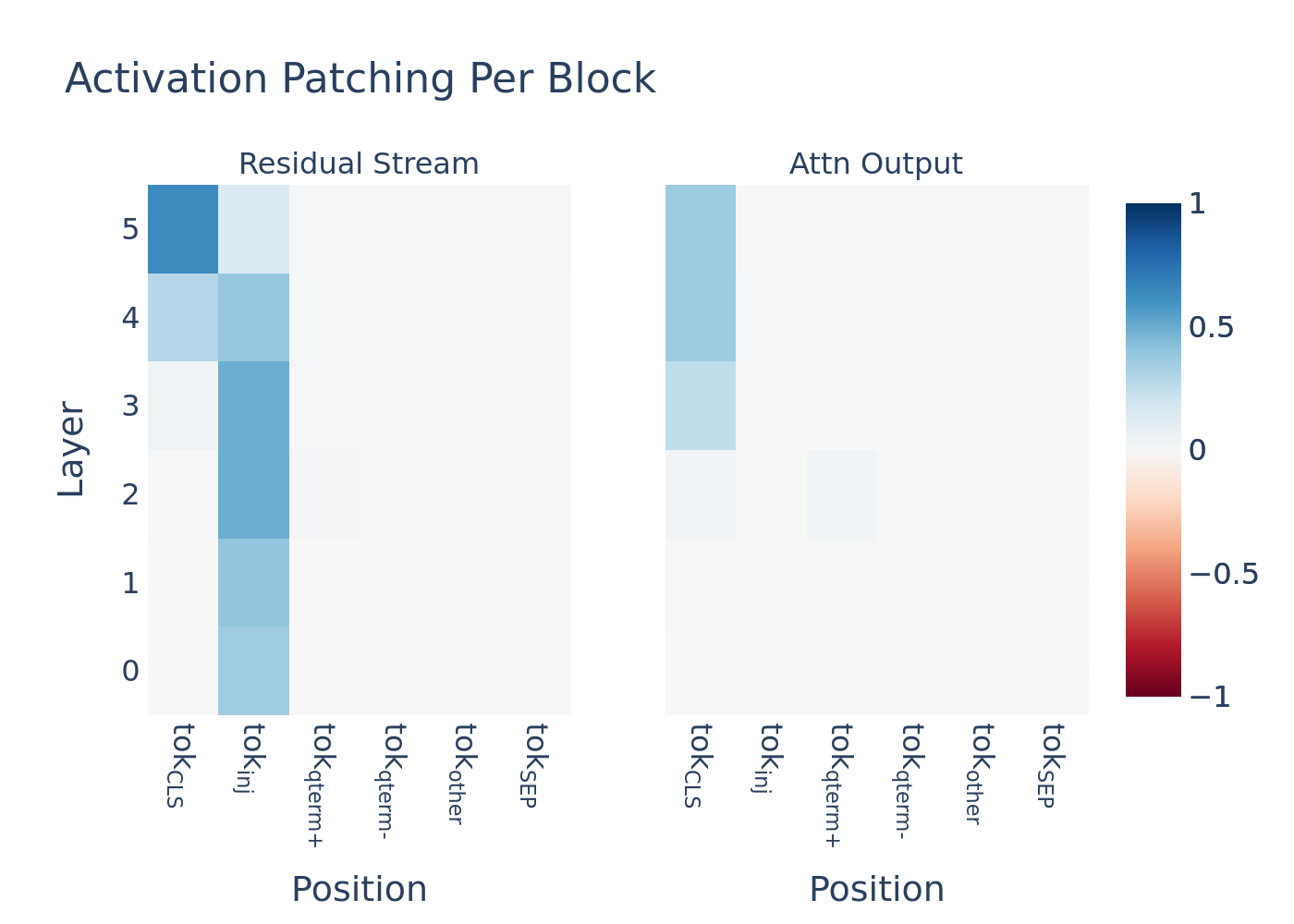}
    \caption{The reproduced block experiment results for prepend from patching into the residual stream and attention output.}
    \label{fig:reproduced_block_prepend}
\end{figure}

 Figure \ref{fig:append_all_block_seg} shows the reproduced results of the append block experiment, while Figure \ref{fig:reproduced_block_prepend} shows the reproduced results of the prepend block experiment. For comparison, Figure \ref{fig:block-append-prepend-original} presents the original results of the block experiment with append and prepend on the residual stream.

 \textbf{Append} Both the original and reproduced heatmaps exhibit strong activations for the injected query term and the pre-existing occurrences of the selected query term in Layer 4, indicating that the existing selected query term occurrences remain relevant in determining the overall performance. Following this, the term frequency information is passed to the CLS token, which shows the strongest activation in Layer 5.

 However, slight variations are observed between the two heatmaps. In Layer 4, the original append heatmap demonstrates a similar impact on the CLS and injected tokens, whereas the reproduced heatmap indicates a stronger impact for the injected tokens compared to the CLS token on the overall performance. Additionally, in Layer 5 of the original heatmap, alongside the strongest impact of the CLS token, a weak impact is visible from the pre-existing occurrences of the selected query term. In contrast, the reproduced heatmap shows this weak impact coming from the injected tokens. Overall, there are slight differences in the impact of some tokens between the original and reproduced heatmaps but they reflect a similar picture.

 \textbf{Prepend} It was suggested by \citet{orig} that the model stores term frequency information in query term positions that are present at the beginning of the document. To verify this, the prepend experiment was done. For this experiment, the heatmaps exhibit a strong impact for the injected query term tokens and show that by Layer 4 the term frequency information is shifted to the CLS token with the strongest impact in Layer 5. However, the original claim about term frequency information being stored at the beginning of the document is not as strongly supported by the updated results, as the strongest activations appear for the injected tokens as opposed to the first appearance of the query term. However, the impact of the injected tokens for the prepend experiment is indeed stronger than for append, and only the append heatmap shows the impact of non-injected query terms as they are closer to the beginning of the document.

Additionally, we have included the heatmaps of the attention outputs for append and prepend that were missing from the original report. For attention outputs, there is a lower impact at later layers at the CLS token position compared to the residual stream. We did not include MLP output results as they do not contain any significant information.

\subsubsection{Activation Patching on Individual Attention Heads}
\begin{figure*}
    \centering
    \includegraphics[width=1\linewidth]{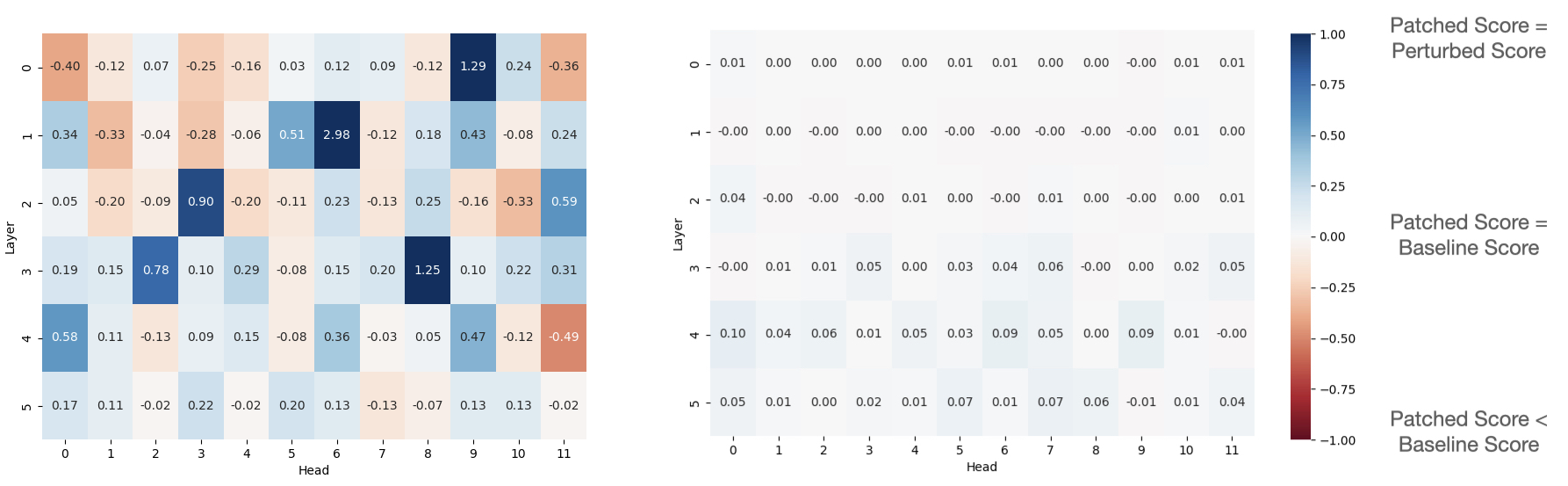}
    \caption{The original results of activation patching on individual attention heads experiment to find the attention heads that encode the TFC1 axiom showing results of the top(left) and bottom(right) 10\% relevant documents per query for append.}
    \label{fig:attention-heads-original}
\end{figure*}

Based on the previous results, \citet{orig} hypothesize that specific attention heads are primarily responsible for processing and carrying the term frequency information throughout the model. To test this hypothesis, activation patching was conducted on the individual attention heads. Figure \ref{fig:attention-heads-original} presents the original results of the attention head experiment with append, showing the results of the top and bottom 10\% relevant documents per query for append. The results with prepend were not mentioned or included in the paper. The results of TFC1-R are mentioned as not relevant as the attention heads amplify an existing relevance signal.

\begin{figure}
    \centering
    \includegraphics[width=1\linewidth]{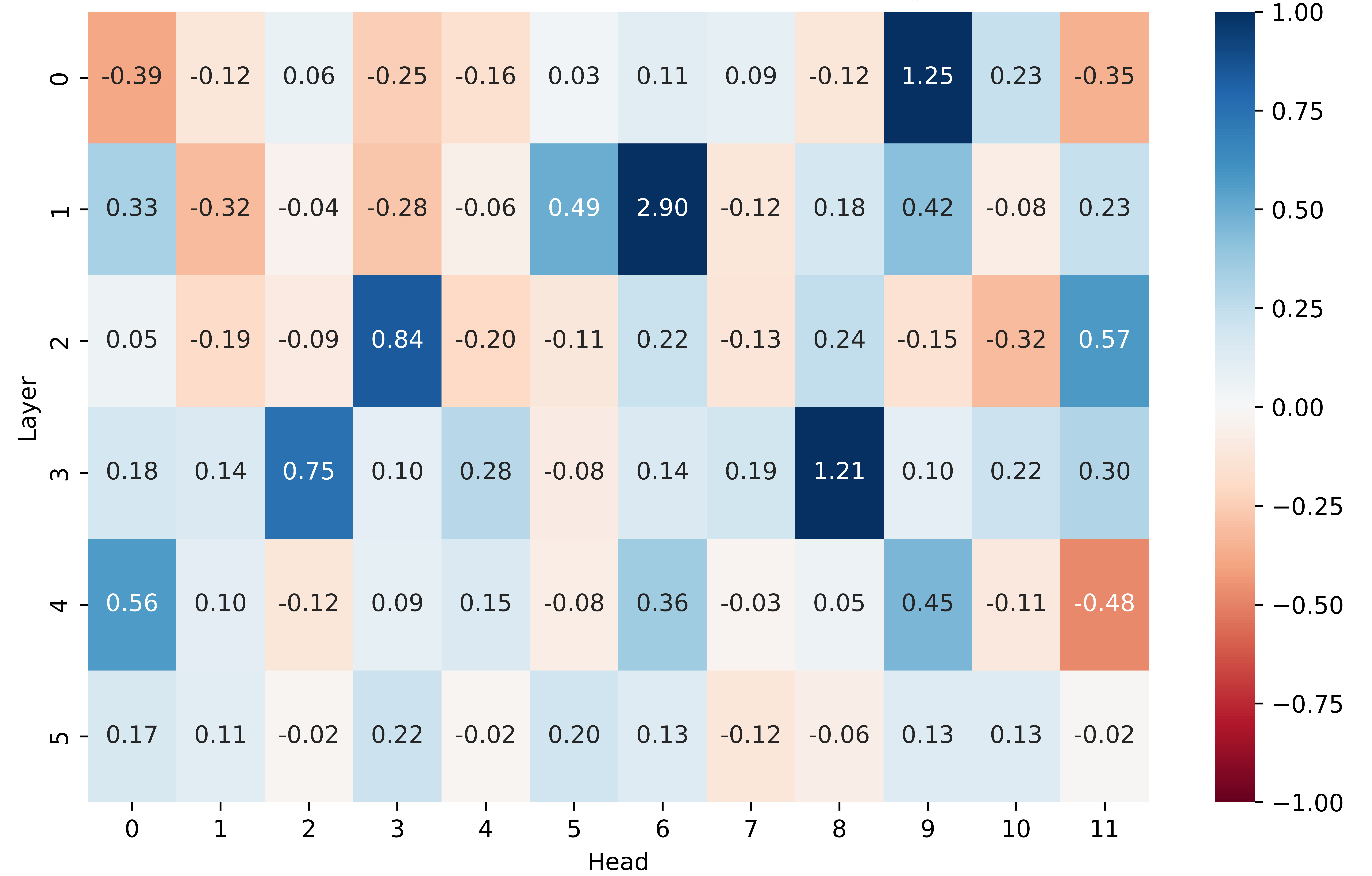}
    \caption{The reproduced results of activation patching on individual attention heads experiment to find the attention heads that encode the TFC1 axiom showing results of the top 10\% relevant documents per query for append.}
    \label{fig:top_ranked_doc_head_results_append}
\end{figure}

\begin{figure}
    \centering
    \includegraphics[width=1\linewidth]{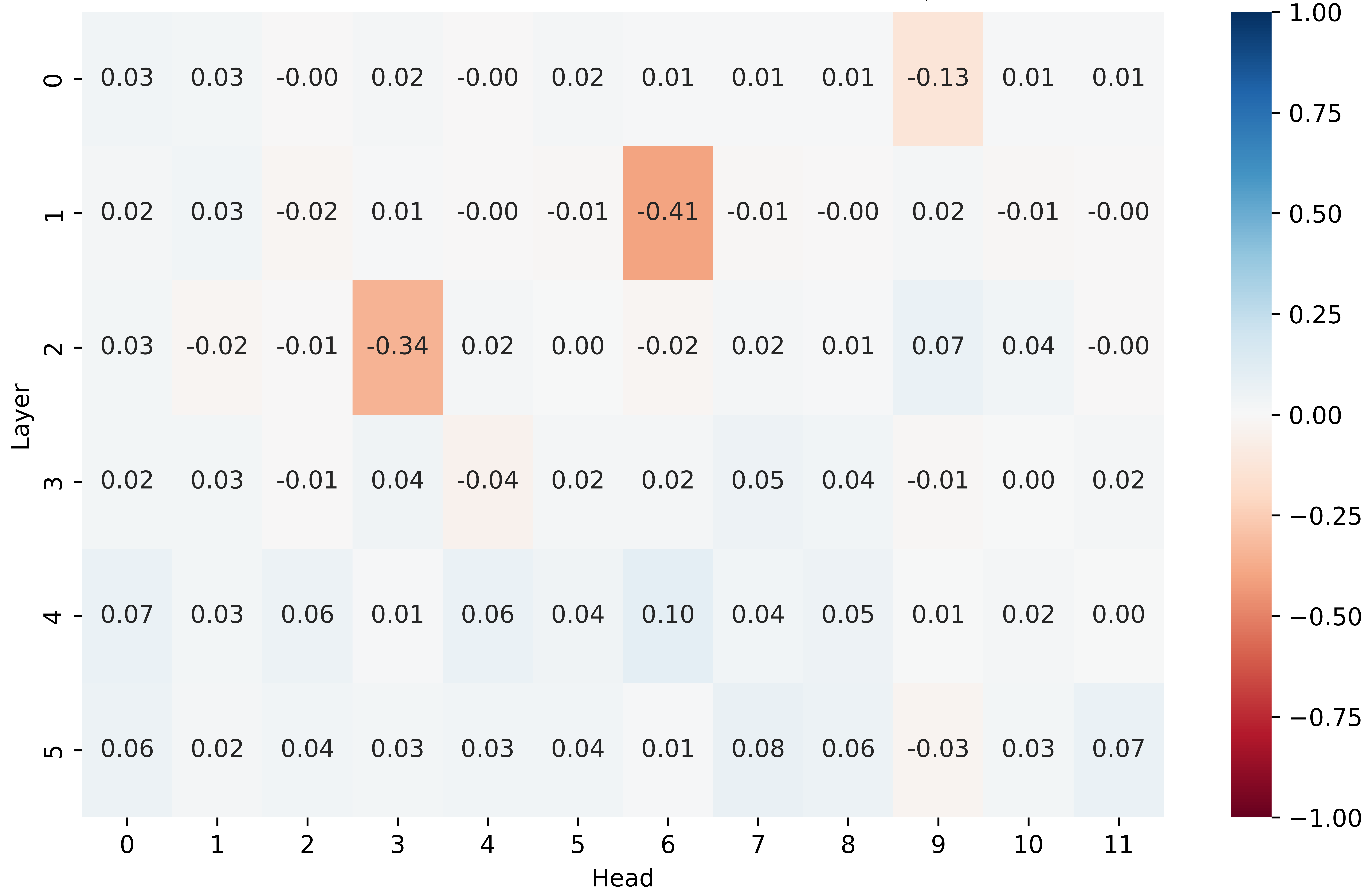}
    \caption{The reproduced results of activation patching on individual attention heads experiment to find the attention heads that encode the TFC1 axiom showing results of the top 10\% relevant documents per query for prepend.}
    \label{fig:top_ranked_doc_head_results_prepend}
\end{figure}

 \textbf{Append} Figure \ref{fig:top_ranked_doc_head_results_append} shows the reproduced results of activation patching on individual attention heads experiment with append for the top 10\% of relevant documents per query. Both the original and reproduced results for the top 10\% might differ slightly in numbers but reveal the same 4 attention heads that encode the TFC1 axiom, i.e. (Layer 0. Head 9), (1.6), (2.3), (3.8). As no significant findings were reported based on the bottom 10 \% of relevant documents, they have been excluded in our paper.
 
 \textbf{Prepend} The results for patching attention head activations with prepend were not reported in the original paper. Based on the results produced by us in Figure \ref{fig:top_ranked_doc_head_results_prepend}, this was because the values across heads are closer to zero for the top-ranked documents, contrary to what we see for append. Similarly, values are close to zero for the bottom-ranked documents as well. Interestingly, the 3 attention heads with the lowest impact were from the list of 4 attention heads with the strongest impact that were identified for the appending perturbation. Therefore, including these results in the appendix should have been done to give more depth to the findings.

\subsection{Extension Results}

\subsubsection{\textbf{LNC1}}

\begin{figure}
    \centering
    \includegraphics[width=1\linewidth]{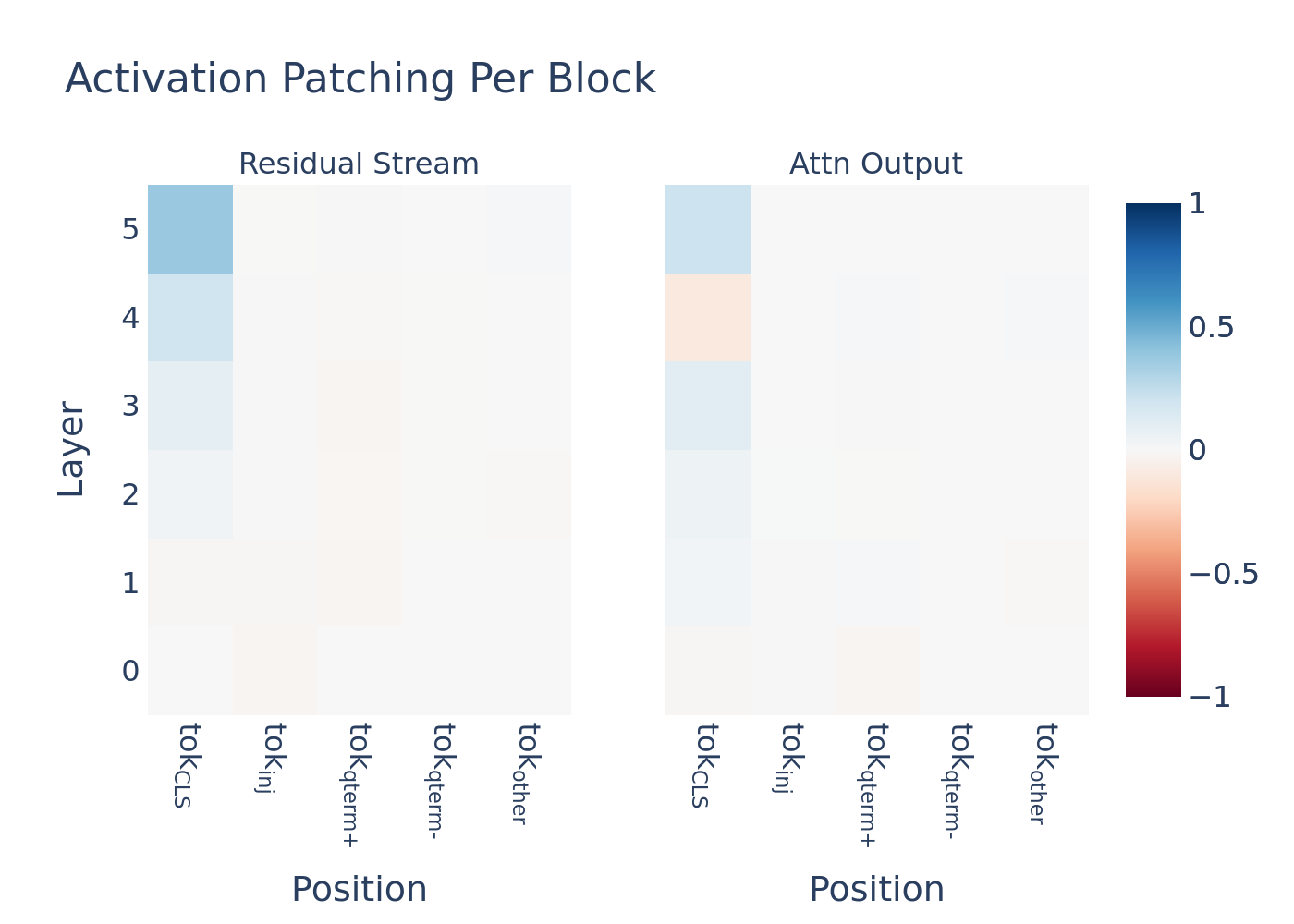}
    \caption{LNC1 block experiment results for append from patching into the residual stream and attention outputs.}
    \label{fig:lnc1_block_append}
\end{figure}

Figure \ref{fig:lnc1_block_append} shows the results of the append block experiment in a heatmap that is similar to the original paper. We report one less column, as in our experiment, we are not injecting query terms. Although this time the patching activation method should decrease relevance, due to the ranking equation, darker blue still represents the most significant impact.

\begin{figure}
    \centering
    \includegraphics[width=1\linewidth]{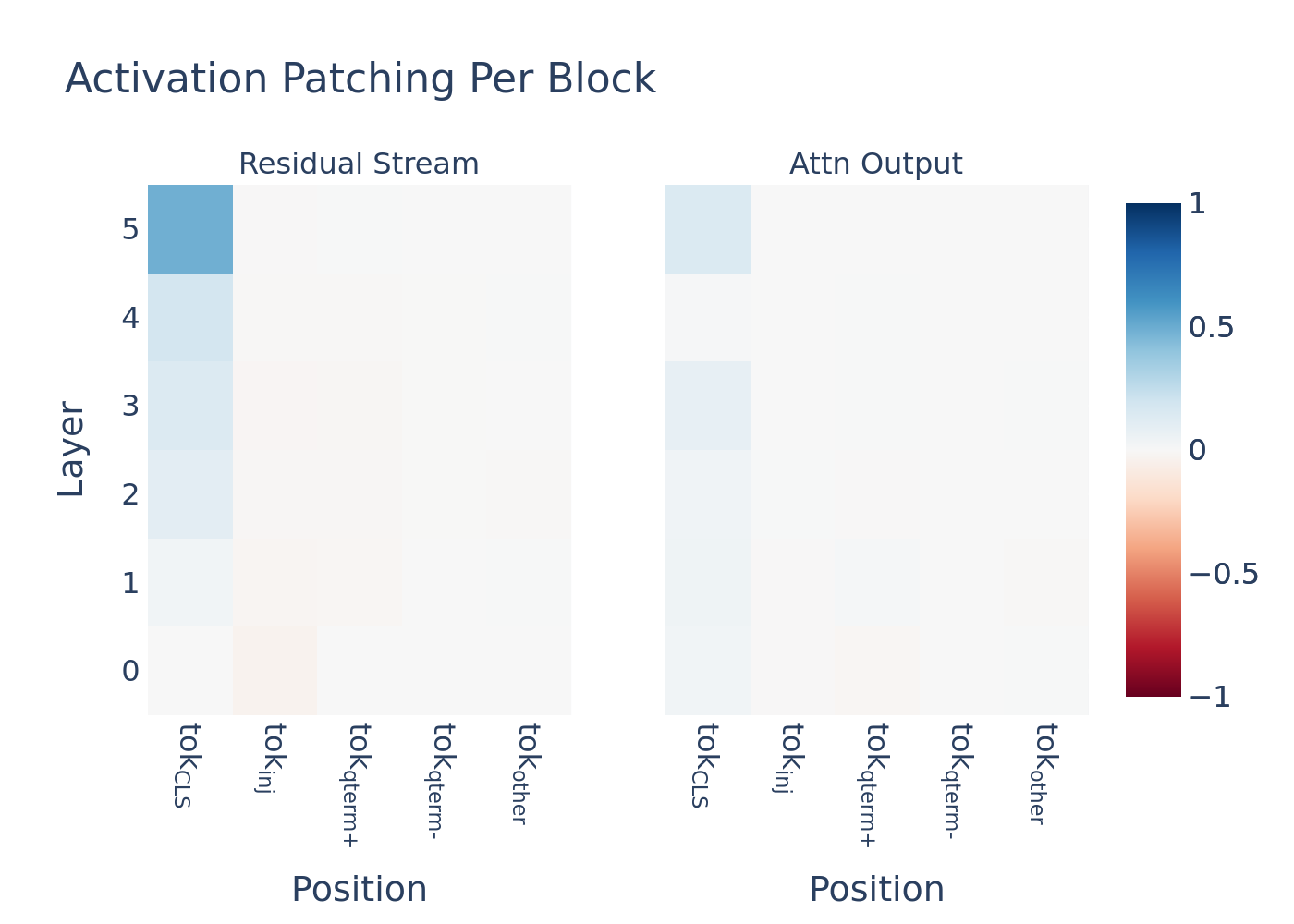}
    \caption{LNC1 block experiment results for prepend from patching into the residual stream and attention outputs.}
    \label{fig:lnc1_block_prepend}
\end{figure}

The append block experiment shows no impact on any token other than the CLS token in the subsequent layers. This is to be expected as document length is not directly connected to any type of token, including the injected ones, as BERT stores sequence-level information in the pooled representation of the CLS token. Furthermore, Figure \ref{fig:lnc1_block_prepend} supports these findings, as injecting tokens at the beginning seems to lead to similar results.

\begin{figure}[ht]
    \centering
    \includegraphics[width=1\linewidth]{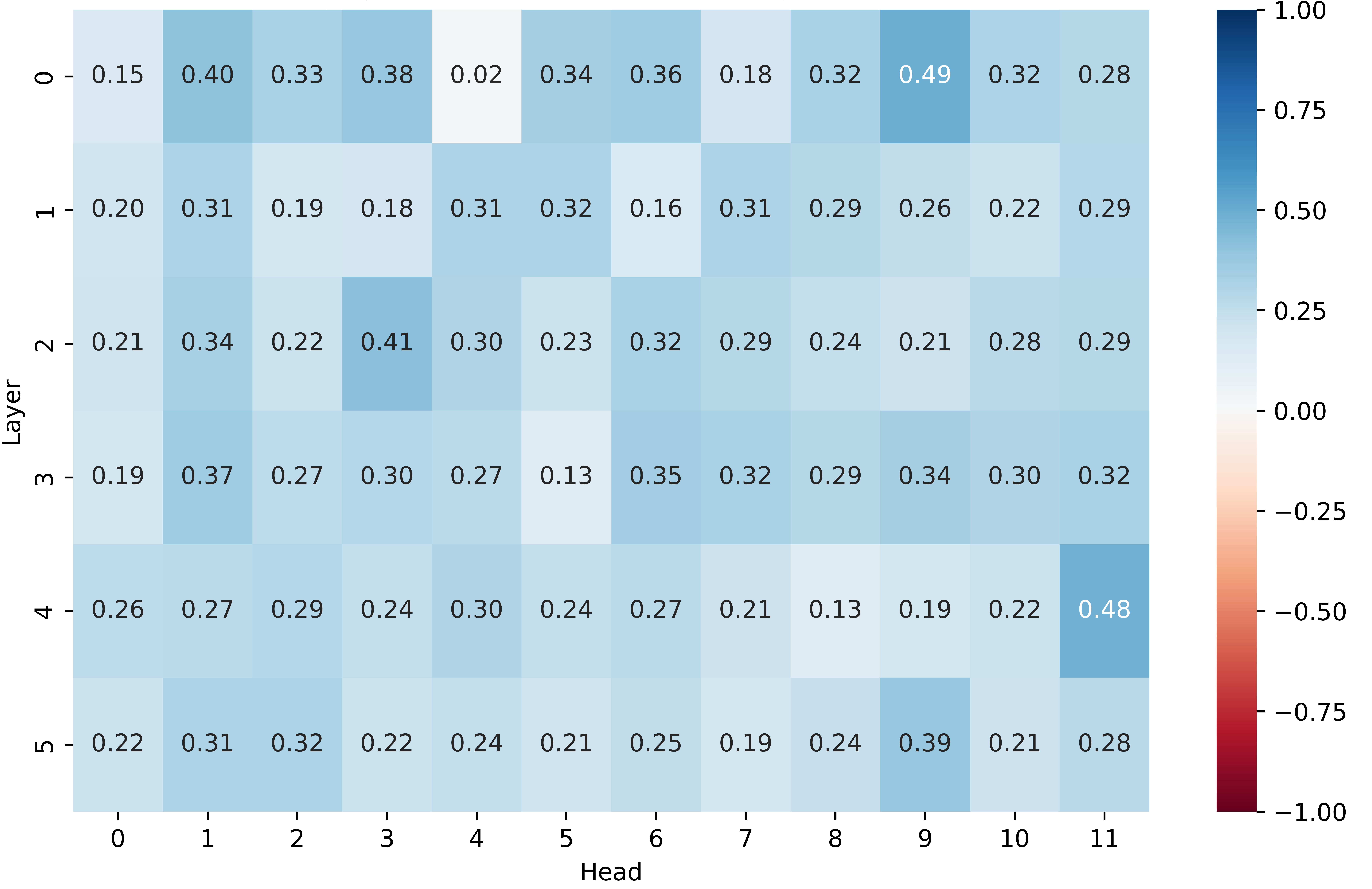}
    \caption{LNC1 results of activation patching on individual attention heads experiment to find the attention heads that encode the TFC1 axiom showing results of the top 10\% relevant documents per query for append.}
    \label{fig:lnc1_top_ranked_doc_head_results_append}
\end{figure}

Figure \ref{fig:lnc1_top_ranked_doc_head_results_append} shows the attention head results of the top 10\% ranked documents for append. The attention head patching has different results from the original paper because, for append, almost all attention heads have a mildly positive impact. 

\begin{figure}[ht]
    \centering
    \includegraphics[width=1\linewidth]{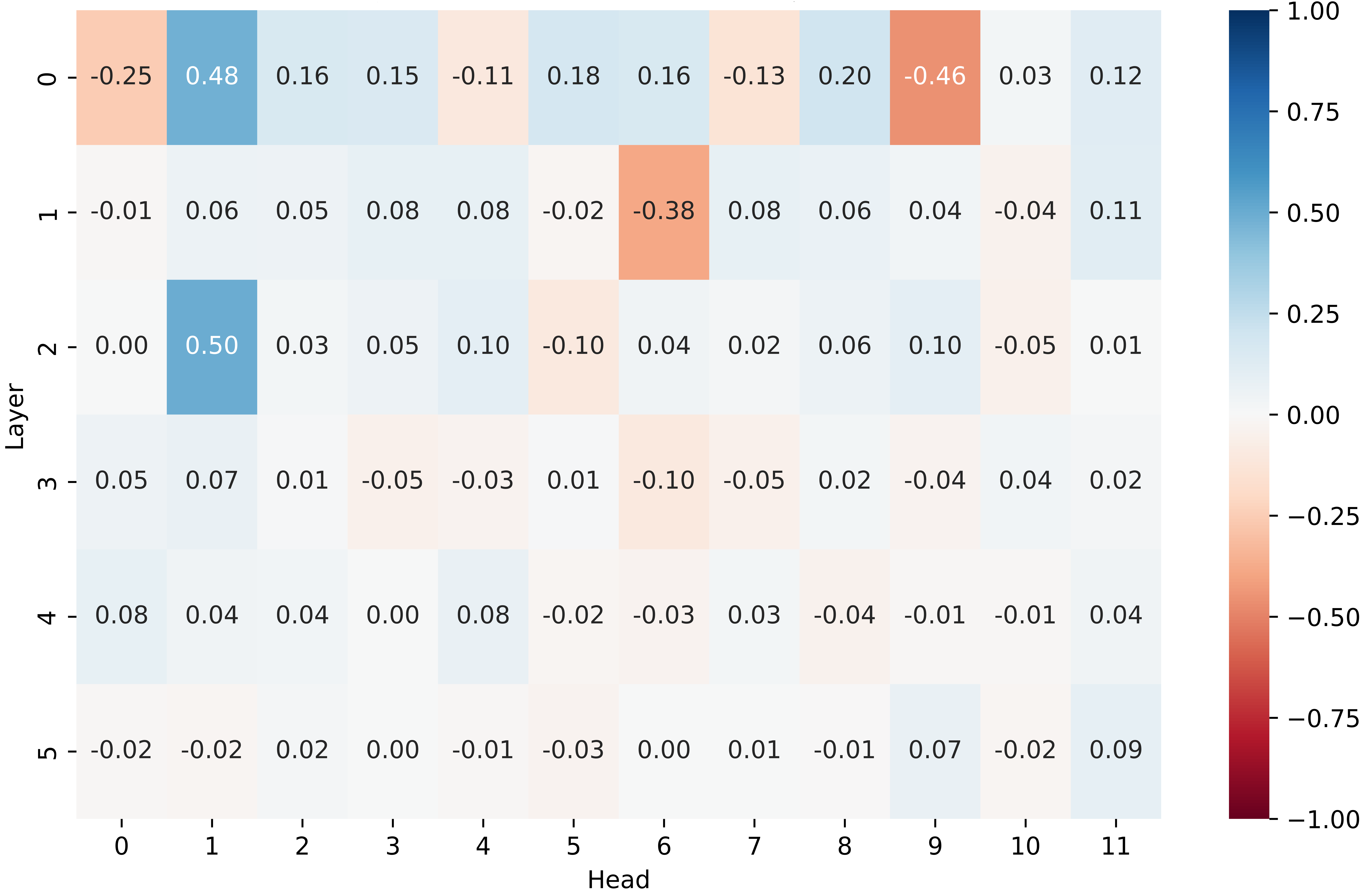}
    \caption{LNC1 results of activation patching on individual attention heads experiment to find the attention heads that encode the TFC1 axiom showing results of the top 10\% relevant documents per query for prepend.}
    \label{fig:lnc1_top_ranked_doc_head_results_prepend}
\end{figure}

Figure \ref{fig:lnc1_top_ranked_doc_head_results_prepend} shows the top 10\% ranked documents for prepend. There are 2 heads that stand out from the rest for their positive impact and 2 that seem to stand out for their negative impact. None of these values are comparable to the values of attention heads identified in the original paper.

\subsubsection{\textbf{mMARCO - Spanish and Chinese}}

\begin{figure}
        \centering
        \includegraphics[width=1\linewidth]{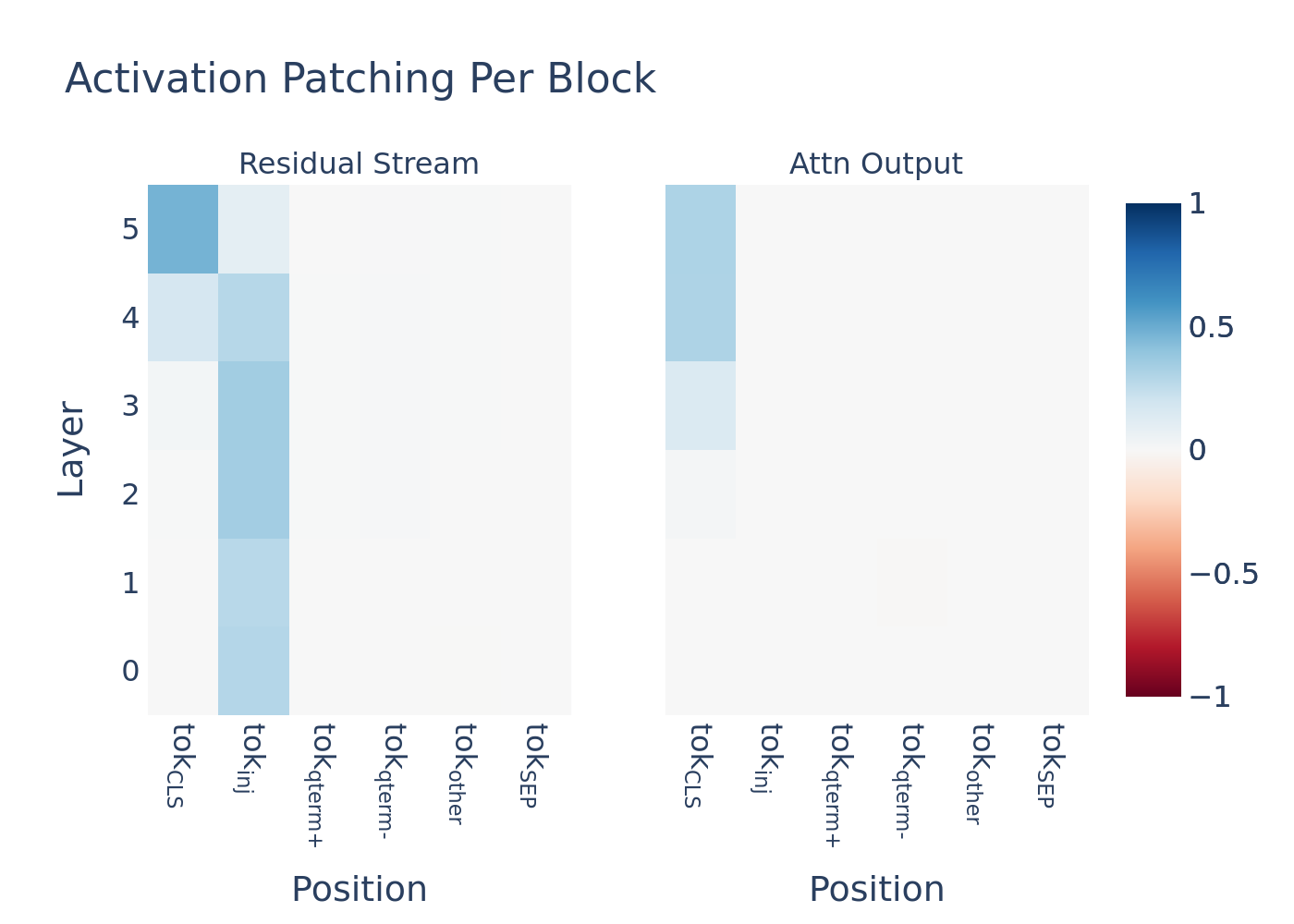}
        \caption{TFC1 mMARCO Spanish language block experiment results for append from patching into the residual stream and attention outputs.}
        \label{fig:es-block-append}
\end{figure}

\begin{figure}
    \centering
    \includegraphics[width=1\linewidth]{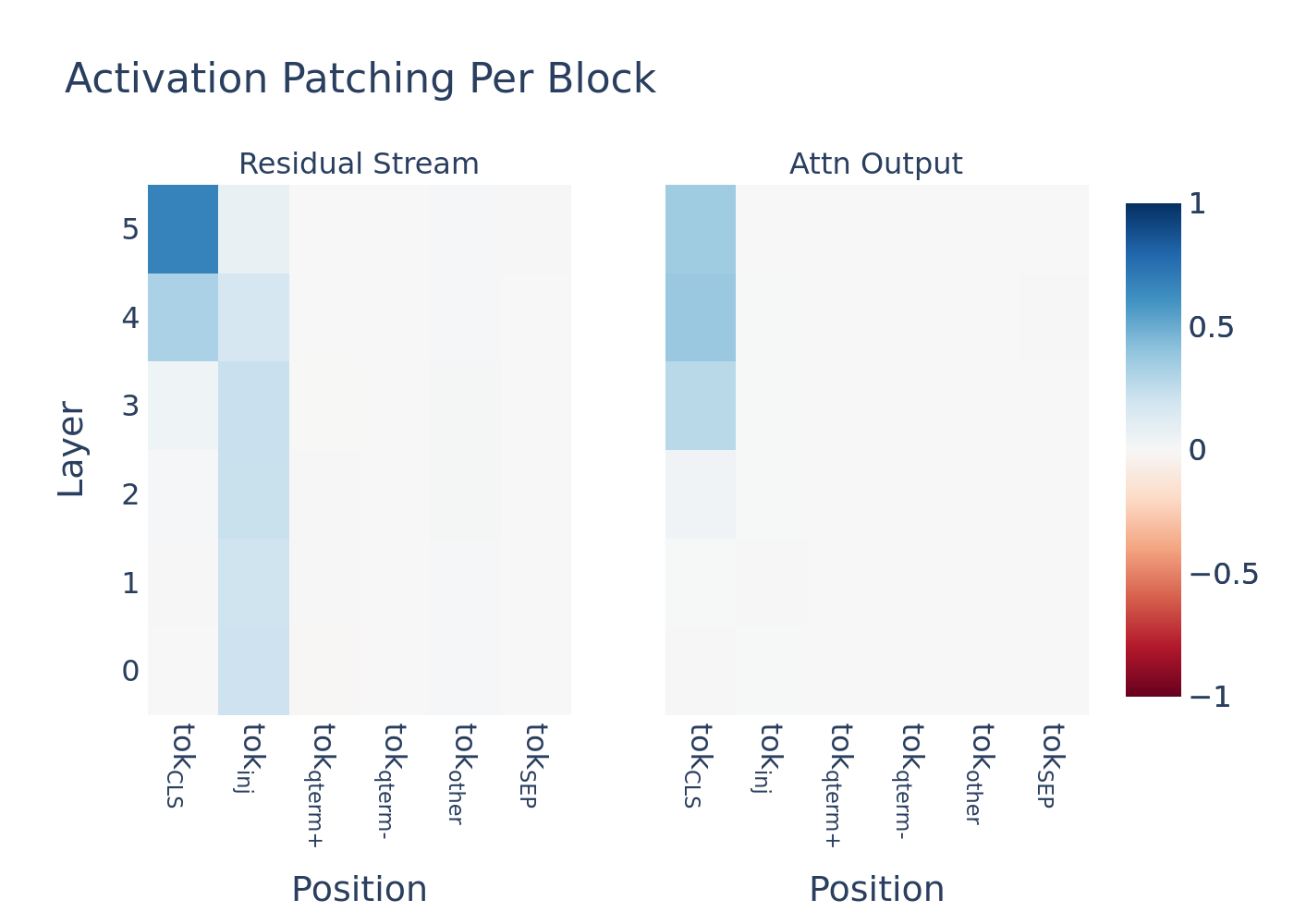}
    \caption{TFC1 mMARCO Chinese language block experiment results for append from patching into the residual stream and attention output.}
    \label{fig:ch-block-append}
\end{figure}
\begin{figure}
    \centering
    \includegraphics[width=1\linewidth]{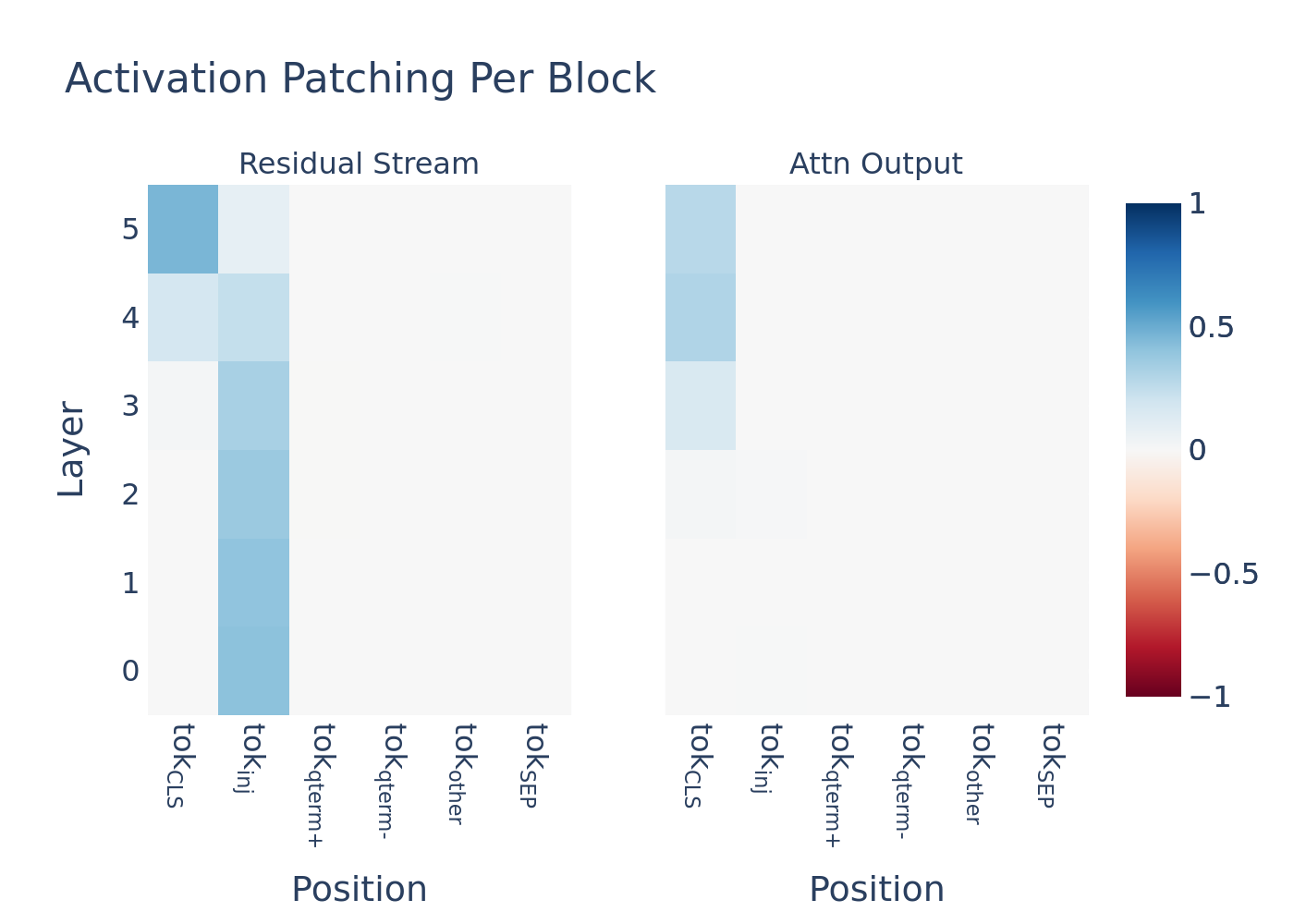}
    \caption{TFC1 mMARCO Spanish language block experiment results for prepend from patching into the residual stream and attention outputs.}
    \label{fig:es-block-prepend}
\end{figure}

\begin{figure}
    \centering
    \includegraphics[width=1\linewidth]{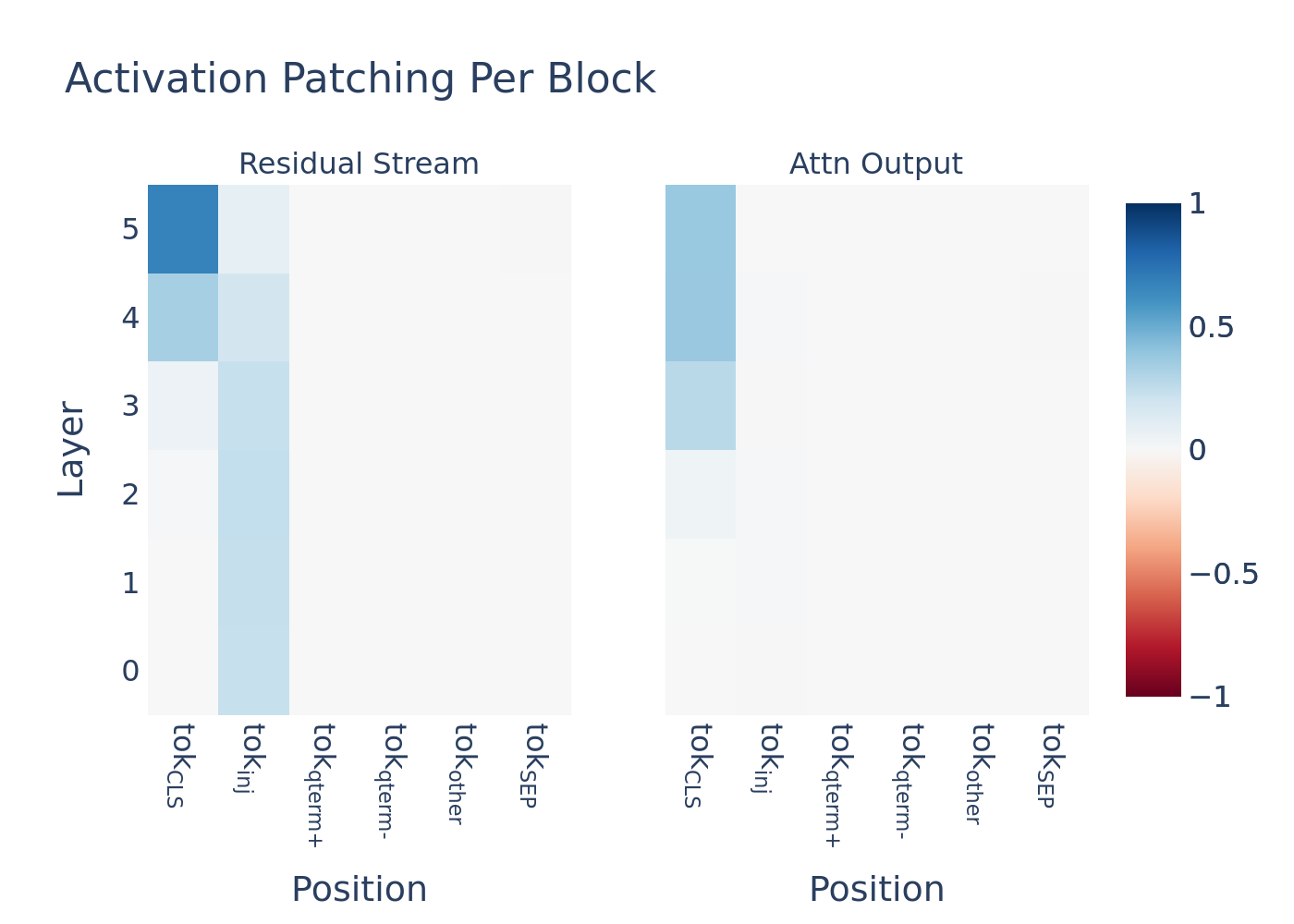}
    \caption{TFC1 mMARCO Chinese language block experiment results for prepend from patching into the residual stream and attention output.}
    \label{fig:ch-block-prepend}
\end{figure}

    
The results of the mMARCO Spanish and Chinese block experiments, shown in Figures \ref{fig:es-block-append} and \ref{fig:ch-block-append}, closely resemble those of the original English experiment. However, the values for injected and the CLS token appear to have a slightly lower impact on ranking for Spanish, while other occurrences of the query term show minimal importance. The prepend results in  Figure \ref{fig:es-block-prepend} and \ref{fig:ch-block-prepend} look similar to the original experiments. Therefore, this extension supports the claim that term frequency is stored across occurrences of the query term, across languages as well. The impact of the injected tokens is slightly higher for prepend, which also shows some support for the information mostly being localized near the beginning of the document.

\begin{figure}[ht]
    \centering
    \includegraphics[width=1\linewidth]{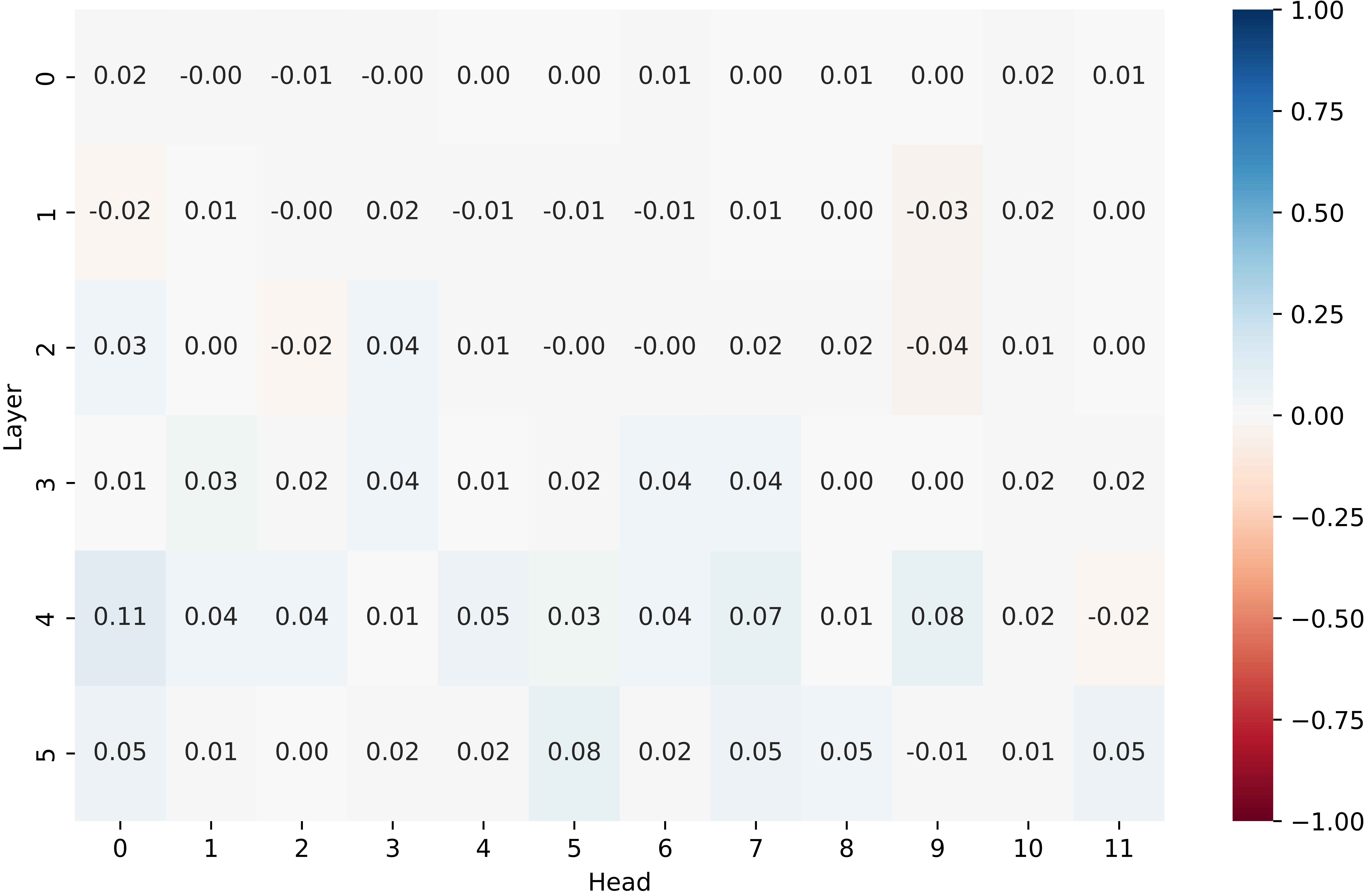}
    \caption{TFC1 mMARCO Spanish language results of activation patching on individual attention heads experiment to find the attention heads that encode the TFC1 axiom showing results of the top 10\% relevant documents per query for append.}
    \label{fig:es-head-append}
\end{figure}

\begin{figure}
    \centering
    \includegraphics[width=1\linewidth]{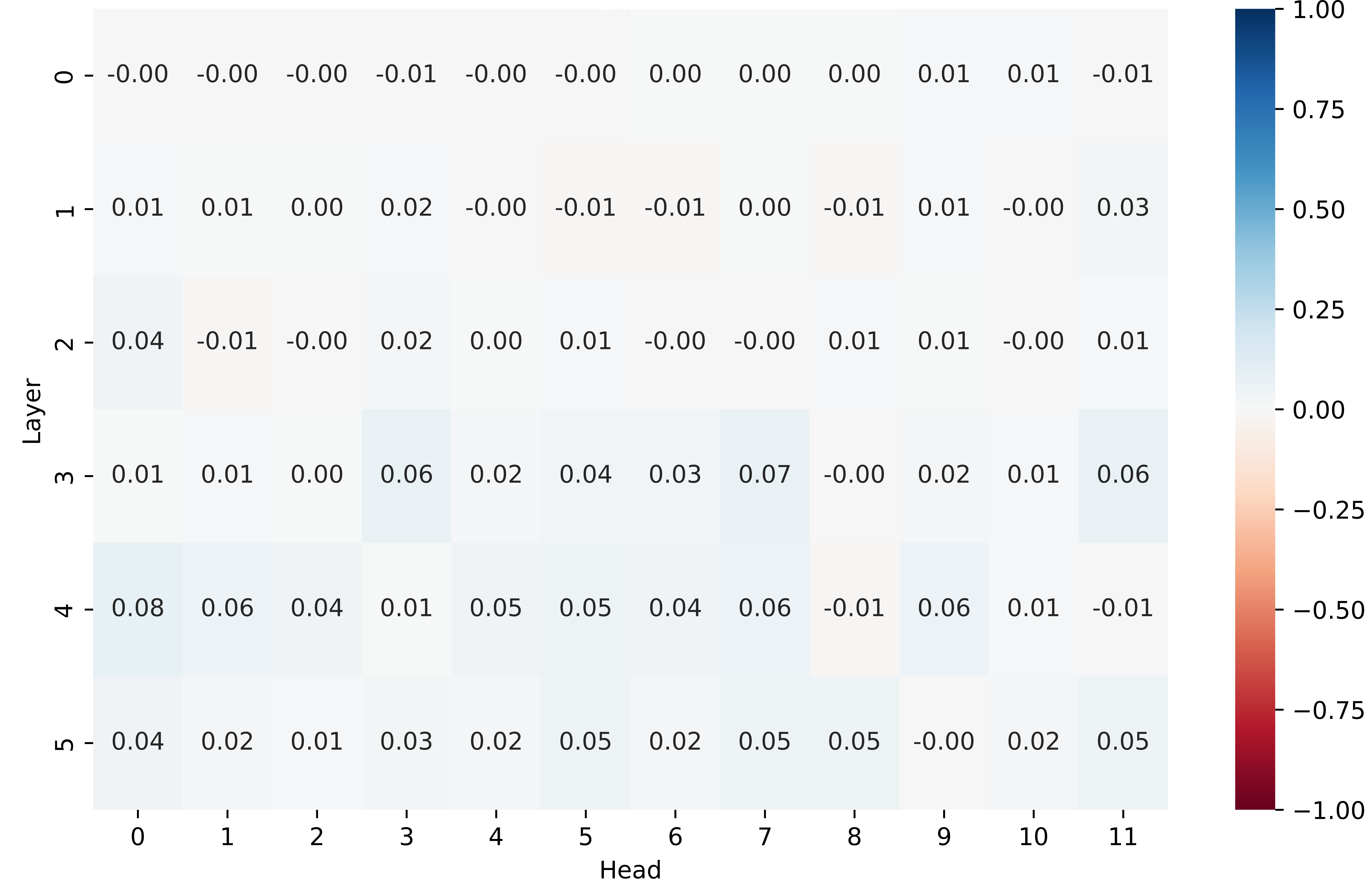}
    \caption{TFC1 mMARCO Chinese language results of activation patching on individual attention heads experiment to find the attention heads that encode the TFC1 axiom showing results of the top 10\% relevant documents per query for append.}
    \label{fig:ch-head-append}
\end{figure}

Figures \ref{fig:es-head-append} and \ref{fig:ch-head-append} show the results of patching attention heads for the top 10\% ranked documents. Compared to the original paper, these results are different because for both prepend and append, the impact of each head seems to be very small and no high-impact heads can be identified. This might be because the model was pre-trained on the English language so there aren't clear attention patterns developed by this model for other languages.

\section{Discussion}
\label{sec:discussion}

The original paper "Axiomatic Causal Interventions for Reverse Engineering Relevance Computation in Neural Retrieval Models" \cite{orig} investigated the internal mechanisms of a transformer-based neural retrieval model. It introduces a novel dataset and activation patching variation suitable for information retrieval. The results of their analysis show that query term frequency information is stored in specific tokens, while at different layers, a few key attention heads play a crucial role in encoding this information. In certain experiments, particularly those involving TFC1-I appending, these tokens and heads are strongly identifiable, demonstrating the effectiveness of the applied method.

Our extension testing of the LNC1 axiom showed that, similar to the TFC1 axiom, the information about document length is stored at the CLS token in the later layers of the model. Additionally, experiments using  Spanish and Chinese datasets confirmed that term frequency information is stored across term occurrences, though the evidence for a stronger signal near the beginning of the document was only marginal. At the same time, neither extension leads to identifying specific attention heads, which is especially unexpected for the Spanish and Chinese language data where the perturbations made are exactly the same kind. This could mean that attention in this model works differently for other languages compared to English. While \cite{orig} included an ablation test to further prove the claim, the fact that these results were only achieved by appending a query term for the top-ranked documents, and not for any other type of perturbation, means that the value of these attention heads should be studied more to get a better understanding of the model.

During the creation of datasets for the extension experiments, we observed the importance of carefully selecting queries and documents where the model correctly increases or decreases the relevance score after axiom-based perturbations. This is due to the fact that the initial activation patching setup requires the model to almost always follow the rules coming from the axiom. However, this is not the case with both the TFC1 and LNC1 axioms.

Additionally, even if we assume the model takes these properties into account, unexpected cases are not accounted for in the design of the ranking metric. It always assumes that the perturbation has the same kind of impact, which in the reported results of the original paper, should always be increasing the relevance score. However, in cases where the baseline score is higher than the relevance score after perturbation, the patching score is most likely to fall between the baseline and perturbed score if the patched module has an impact. This issue became apparent when we initially ran extension experiments on datasets that did not consistently produce axiom-compliant behaviour. As a result, some findings contained a mix of positive and negative scores, making interpretation difficult. Addressing these cases requires further investigation, and potentially, a more adaptive ranking metric should be developed to capture variations more accurately.
\section{Conclusion}
\label{sec:conclusion}

In this paper we reproduce and extend “Axiomatic Causal Interventions for Reverse Engineering Relevance Computation in Neural Retrieval Models”, aiming to validate their original findings. We successfully reproduced most of the key experiments, revealing that the TAS-B model indeed stores term frequency information across occurrences of the query term, possibly more at the start of the document. Additionally, we could see that the neural retrieval model moves this information to the CLS token pooled representation in the later layers. Our LNC1-based experiments confirmed the importance of the CLS token for encoding document-level features while our extension with the Spanish and Chinese mMARCO datasets showed that some of these term-frequency mechanisms generalize beyond English.

In summary, this work reaffirms that the designed activation patching method can uncover specific mechanisms of relevance modelling, such as term frequency and document length encoding. However, it also shows the need for more robust interpretability approaches and more focus to guarantee the reproducibility of machine learning research. Future research could focus on investigating circuits of components instead of singular modules to obtain a clearer picture of how information is processed while also further looking into some of the findings of this paper. This could include testing different ranking metrics and perturbations to gain a wider knowledge of the inner workings of neural retrieval models. Perhaps, that will assist in identifying specific mechanisms across other tasks and languages.

\newpage
\newpage
\bibliographystyle{ACM-Reference-Format}
\balance

\appendix

\end{document}